\RequirePackage{fix-cm}
\documentclass{svjour3}                     
\smartqed  
\usepackage{graphicx}
\usepackage{xcolor}
\usepackage{comment}
\usepackage{hyperref}
\usepackage{lineno}
\usepackage{ulem}
\newcommand{\Add}[1]{\textcolor{black}{#1}}
\newcommand{\Erase}[1]{\if0{#1}\fi}

\begin{document}
\abovedisplayskip=2pt
\belowdisplayskip=2pt

\title{Broadband multi-layer anti-reflection coatings with mullite and duroid \Erase{used}for half\Add{-}wave plate\Add{s} and alumina filter\Add{s} for CMB polarimetry 
}


\author{Kana Sakaguri${}^{1}$ \and Masaya Hasegawa${}^{2}$ \and  Yuki Sakurai${}^{3}$   \and Charles Hill${}^{4,5}$  \and Akito Kusaka${}^{1,3,4}$   
}



\institute{${}^{1}$Department of Physics, University of Tokyo, Bunkyo-ku, Tokyo 113-0033, Japan\\
              \email{ksakaguri@cmb.phys.s.u-tokyo.ac.jp}           \\
           ${}^{2}$High Energy Accelerator Research Organization, Tsukuba, Ibaraki 305-0801, Japan\\
           ${}^{3}$Kavli IPMU, University of Tokyo, Kashiwa, Chiba 277-8583, Japan\\
           ${}^{4}$Physics Division, Lawrence Berkeley National Laboratory, Berkeley, CA 94720, USA\\
           ${}^{5}$Department of Physics, University of California, Berkeley, CA 94720, USA
}

\date{Received: date / Accepted: date}

\maketitle

\begin{abstract}

\Erase{We developed a}\Add{A} broadband two-layer anti-reflection (AR) coating \Add{was developed} for use on a sapphire half-wave plate (HWP) and an alumina infrared (IR) filter for cosmic microwave background (CMB) polarimetry. Measuring tiny CMB B-mode signals requires maximizing the number of photons reaching the detectors and minimizing spurious polarization due to reflection with an off-axis incident angle. However, a sapphire HWP and an alumina IR filter have high refractive indices of $\simeq$ 3.1, and an AR coating must be applied to them.
Thermally sprayed mullite and Duroid~5880LZ were selected in terms of index and coefficient of thermal expansion for use at cryogenic temperatures. With these materials, \Erase{we reduced}the reflectivity \Add{was reduced} to about 2\% at 90/150~GHz and $<$1\% at 220/280~GHz. \Erase{We describe the}\Add{The} design, fabrication, and optical performance evaluation of the AR coatings \Add{are described}. The coatings were used in a current ground-based CMB experiment called the Simons Array. They could also be applied to next-generation CMB experiments\Add{, such as} \Erase{like}the Simons Observatory.
\end{abstract}

\section{Introduction}
\label{intro}

\Add{The} cosmic microwave background (CMB) has a variety of information that is useful for understanding the early universe \cite{PhysRevD.60.043504,PhysRevLett.78.2058}. In particular, B-mode polarization is a unique pattern of parity-odd polarization that derives from primordial gravitational waves and gravitational lensing. The observation of B-mode polarization from primordial gravitational waves would provide strong evidence of inflation.

For high-precision CMB experiments, \Erase{there have been}recent advances in the development of devices to reduce systematic errors originating from optical systems and to improve sensitivity \Add{have occurred}. Optical elements\Add{,} such as \Erase{a}continuously rotating half-wave plate\Add{s} (HWPs) and \Erase{a}filter\Add{s,} have been introduced. A\Erase{n} HWP modulates the polarization signal, and the filter removes infrared (IR) \Add{signals}. However, the sapphire and alumina used for HWPs and IR filters have high refractive indices of $\simeq$ 3.1 \cite{Inoue:16,Lamb}. Thus, they refract much light and increase systematic and statistical uncertainties. 

\Erase{In order to}\Add{To} solve this problem, an anti-reflection (AR) coating is \Erase{particularly}critical for these materials. \Add{Some ways to apply AR coatings like layering dielectrics or machining sub-wavelength structures have been reported \cite{Raut2011,Takaku_2021}. In these methods, layering dielectrics were chosen in terms of the hardness of machining substrates, which are sapphire or alumina, machining speed of large-diameter substrates, and application possibility to the high-frequency band.} Our AR coating consists of two dielectric layers and can reduce reflectivity using the materials with different refractive indices on the surface of optical elements. By adjusting the thickness of the coating materials, an AR coating can be optimized for a specific frequency band.

In this paper, \Erase{we present}the development of broadband multi-layer AR coatings for CMB experiments with 90/150~GHz and 220/280~GHz dichroic detectors that each cover approximately 30\% of the fractional bandwidth \Add{is presented}. To realize this broadband frequency coverage, \Erase{we developed}two-layer coatings with an average reflectivity of less than 3\% \Add{were developed}. \Erase{There are}Two challenges regarding AR coatings for CMB observations \Add{should be mentioned}. 
First, it is difficult to apply a coating on 50-cm-diameter sapphire or alumina with uniformity of tens of microns.
Second, the AR-coated optical elements are cooled down to 40~\Erase{K}or 4~K \Erase{in order}to reduce thermal radiation on the cryogenic detectors, which creates challenges associated with differential thermal contraction.

For the first layer, thermally sprayed mullite \cite{Inoue:16}, a ceramic material (Tocalo Corporation) \cite{TOCALO}, was used, while the second layer was Duroid 5880LZ, a composite material (Rogers Corporation) \cite{ROGERS}. These materials were selected both for their refractive index and coefficient of thermal expansion. \Erase{We discuss the}AR coating's design, fabrication, and its optical performance at room temperature \Add{are discussed}.

\section{Design and fabrication}
\label{sec:design and fab}

\begin{figure}[tbp]
\centering
  \includegraphics[width=10 cm, clip]{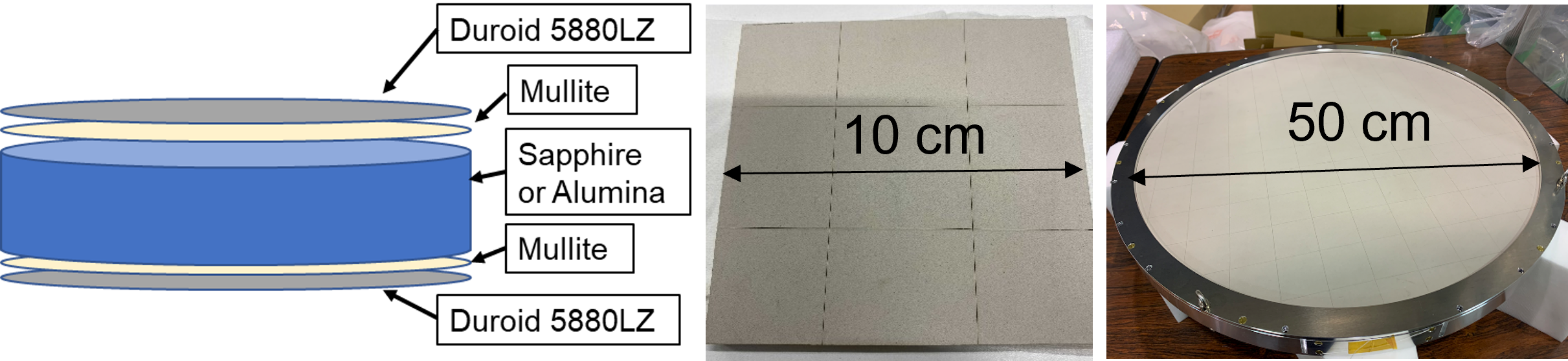}
\caption{Left: Schematic view of anti-reflection (AR) coatings. The AR coatings consist of the thermally-sprayed mullite as the first layer and Duroid 5880LZ glued with Epo-Tek as the second layer. Middle: Small test sample with AR coatings. This small sample was fabricated to check that it would not delaminate and also to check its optical performances. Right: Simons Array HWP with AR coatings for 90/150~GHz.}
\label{fig:AR conceptual diagram}
\end{figure}

The target\Add{s} of our AR coating \Erase{is}\Add{are} ground-based CMB experiments, and the requirements of the HWP and IR filter are \Add{described bellow}\Erase{as follows}:
\begin{itemize}
    \item The diameters of the HWP and the filter are \Erase{as large as}about 50~cm, and the AR coating needs to be applied evenly to this large diameter \cite{Hill:SPIE,Ali_2020}.
    \item The reflectance should be reduced to a few percent at the detector bands.
    Roughly, 80-110/130-170~GHz and 200-260/260-320~GHz were chosen to calculate the average reflectance for each one \cite{Abitbol2020,10.1117/12.2312821,abazajian2019cmbs4}.
    \item The AR coatings should not delaminate when cooled because the filter and the HWP are used at 40-50 K, and $\sim$4 K in some cases \cite{Ali_2020}.
\end{itemize}
\Add{Multiple reports have been made so far \cite{Golec2020,Nadolski2018,Rosen:13}, but producing repeatable at large diameters is still challenging.}
To achieve such broadband and low-reflectivity coatings, \Erase{we developed}two-layer coatings \Add{were developed in our study}.
\Add{In our coating, the reflectance was minimized at a specific frequency band by layering dielectrics with different refractive indices, which step down to a refractive index of 1 for air. First, the optimal indices of the layers were calculated. Using the condition that the optical thickness of each layer was $\lambda/4$ wavelengths relative to the incident wavelength~$\lambda$, the optimal index relation could be derived:}
\begin{equation}
    \Erase{1 = \frac{n_1^2 n_{\rm{Sapphire\, or\, Alumina}}}{n_2^2},}
    \Add{n_s = \frac{n_2^2}{n_1^2}},
\end{equation}
\Add{for which}\Erase{where} $n_1$, $n_2$, and $n_{\Add{s}}$ are the indices of the first layer, second layer, and sapphire or alumina, respectively. This \Add{order originates}\Erase{comes} from the transmitted and reflected waves being 180 degrees out of phase, and the interference \Add{causes a reduction in} the reflectance \cite{1989itcm.book.....M}.
Considering the index, coefficient of thermal expansion, and compressive modulus, \Erase{we chose}coating materials \Add{were chosen} to satisfy the \Add{study} requirements. \Erase{mentioned.}Figure \ref{fig:AR conceptual diagram} shows a diagram of our AR coatings. \Erase{We selected} Mullite ceramic~\cite{TOCALO} \Add{was selected} as the first layer and Duroid 5880LZ \cite{ROGERS} aluminosilicate microspheres as the second layer.
\Add{Mullite had been used before \cite{Inoue:16}, but the combination of mullite and Duroid was used for the first time.}
The properties of the coating layers are listed in Table \ref{tab:AR properties}.

\begin{table}[tbp]
 \caption{Basic properties of our AR coating materials. The AR indices, $n$, and thicknesses, $d$, for 90/150~GHz and 220/280~GHz coatings are shown. The thicknesses are the fabricated values, and the errors represent the production errors. All values are at room temperature.}
  \centering
   \begin{tabular}{c|c|c|c} \hline
     Material  & $n$    &   $d$~[mm] (90/150~GHz) & $d$~[mm] (220/280~GHz)  \\ \hline
     Mullite   & 2.52 $\pm$ 0.02 \cite{Inoue:16} &   0.254 $\pm$ 0.01     & 0.147 $\pm$ 0.01                \\ \hline
     Duroid    & 1.41 $\pm$ 0.01 \cite{ROGERS} &   0.385 $\pm$ 0.01              & 0.155 $\pm$ 0.01                \\ \hline
   \end{tabular}
  \label{tab:AR properties}
\end{table}

After selecting the coating materials, small test samples were fabricated after calculating the thicknesses of layers by minimizing the average reflectance between the frequency bands.
\Add{The predicted reflectance was calculated as an infinite plane wave incident on a homogeneous dielectric thin layer.}

The middle part of Figure \ref{fig:AR conceptual diagram} shows a photo of the fabricated AR coating sample. \Erase{We used}A 10-cm square of alumina with a thickness of 4~mm \Add{was used}.
The first layer of mullite was thermally sprayed on both alumina surfaces \cite{Inoue:16} by the Tocalo Corporation.
\Add{This process uses an established and available technology and can be fabricated to a high precision of 10~{\textmu}m.}
\Add{Because mullite has a coefficient of thermal expansion matched to that of alumina and adheres better to alumina, it would not cause delamination under cryogenic conditions \cite{Inoue:16}.}
The second layer of Duroid was glued on with \Add{40-{\textmu}m-thick} Epo-Tek, a type of epoxy.
\Add{Duroid thickness was machined to 385$\pm$10~{\textmu}m at Suzuno Giken \cite{Suzuno}.}
\Add{To compute the accurate model during simulation, the reflectance was effectively calculated with three layers including the thin layer of Epo-Tek.}
\Add{Thickness of the Epo-Tek layer is 20-40{\textmu}m. While this may appear rather large uncertainty, there is relatively an impact on the transmission due to this thickness because the Epo-Tek layer is thin and its index is almost exactly the mean of the indices of mullite and Duroid.}
\Erase{We checked}The optical performance at 300~K \Add{was checked,} and \Erase{reoptimized}the thickness \Add{was reoptimized} while considering the thickness trend resulting from the optical measurement. This process was repeated until the best coating was finally obtained.

A large-diameter sample was fabricated in the same way as the small sample after determining the optimal thickness with a small test sample, thus completing the development of coatings for 90/150~GHz. A\Erase{n} HWP sandwiched between AR-coated alumina (Figure \ref{fig:AR conceptual diagram}) and an IR filter with our coatings were used in an experiment with the telescope of the Simons Array.

\section{Optical performance and cryogenic validation}
\label{sec:opt.meas.}

\begin{figure}[tbp]
\centering
  \includegraphics[width=9 cm,
  clip]{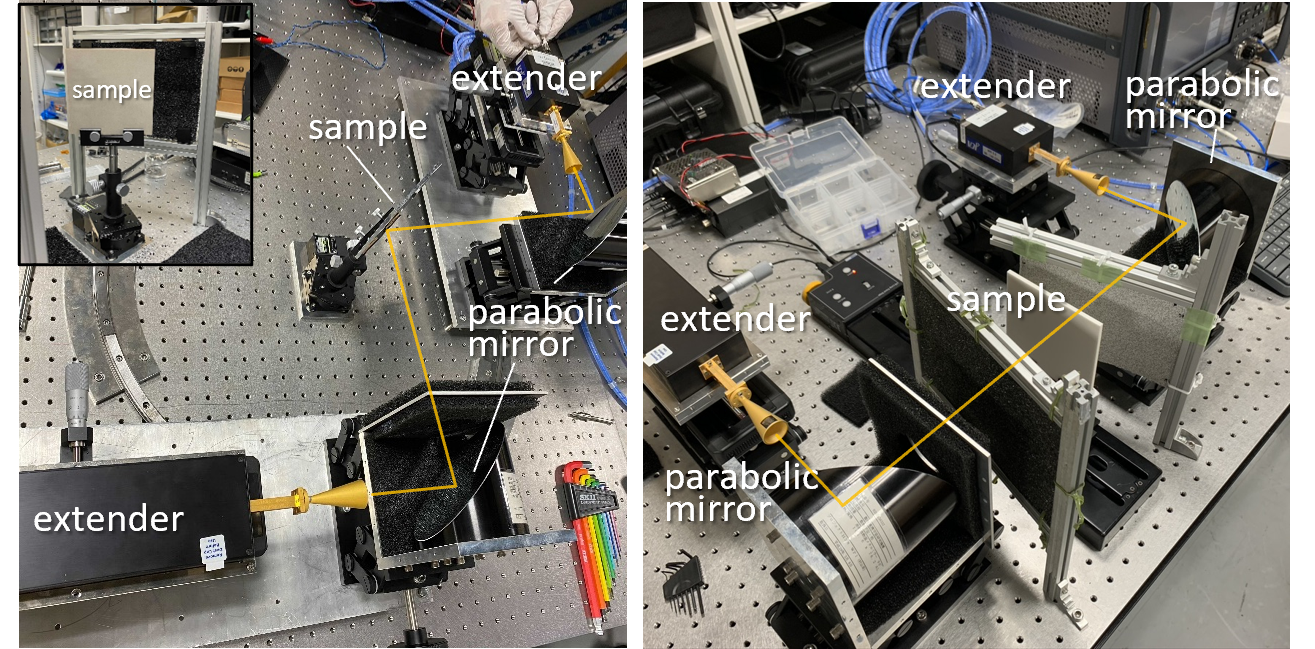}
\caption{Left: Measurement setup for reflection. The yellow lines indicate the optical path. The electromagnetic wave output from the extender, \Add{which is placed at the focal distance of a parabolic mirror,} is reflected by the mirror, \Add{and become a plane wave. It} enters the sample as s-polarized light. The light reflected through the sample enters the extender on the receiving side. The top left photo shows the vertical view of the sample. Right: Setup for transmission.}
\label{fig:setup roomtemp}
\end{figure}

The optical performance of the small samples was analyzed using a vector network analyzer (VNA) at Kavli Institute for the Physics and Mathematics of the Universe (IPMU). The reflectivity at an incident angle of 45 degrees and transmissivity of 55 to 330~GHz was measured, which is the required frequency range for CMB observation. The measurement setup at room temperature is shown in Figure~\ref{fig:setup roomtemp}.
\Add{We measured an alumina slab without AR before measuring AR coating samples as a validation of the setup and confirmed that the same fringe patterns were measured as predicted.}

\begin{figure}[tbp]
\begin{minipage}[t]{0.5\linewidth}
\centering
  \includegraphics[keepaspectratio,
  scale=0.18]{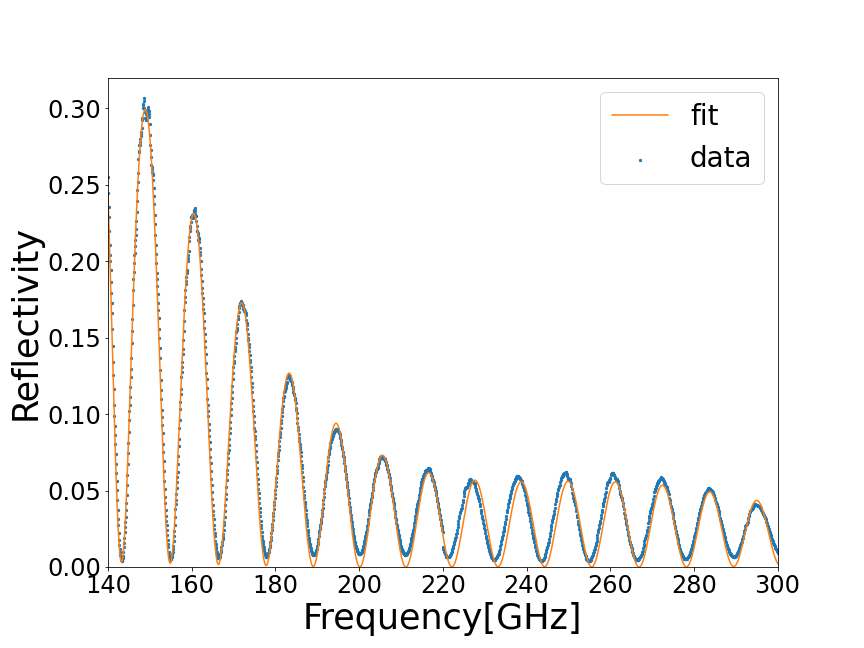}
\label{fig:ref fit UHF}
\end{minipage}
\begin{minipage}[t]{0.5\linewidth}
\centering
  \includegraphics[keepaspectratio,
  scale=0.18]{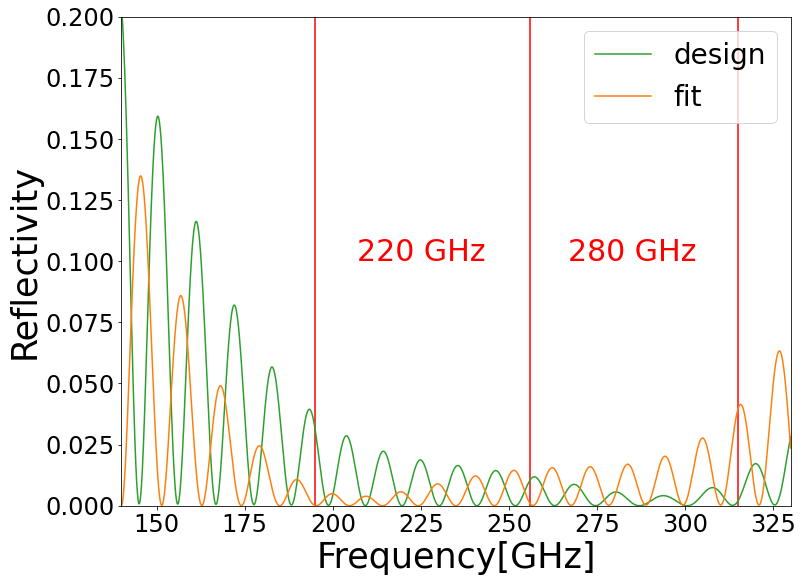}
\label{fig:ref UHF onaxis}
\end{minipage}
\caption{Left: Measured data between 140 and 330~GHz and fitting of the data using analytic simulation.
Right: On-axis performance using the analytic fit compared to the design values. Red vertical lines show the frequency band range that we calculated the average reflectance.}
\label{fig:ref result}
\end{figure}

Figure \ref{fig:ref result} shows the measured reflectivity (an example from the 220/280~GHz sample).
\Add{First, measured reflection at 45-degree incidence was fitted with thickness and index of each layer.
Optical measurements were performed once after spraying mullite but before applying Duroid coating, so the resulting thickness and index were reliable. The fit values were then substituted in the model with a zero-degree incident angle. Finally, the average reflectance was calculated at zero degrees at the detector bands.}
Some samples were also measured at a different incidence angle (18~degrees) to confirm that this analytical process was accurate.

Table \ref{tab:relfectivity} shows the average reflectance results of the small samples. It was possible to make samples with an average reflectance of $<$3\% for 90/150~GHz and 1\% for 220/280~GHz. The difference between the sample and \Erase{the}design mainly \Add{originates}\Erase{comes} from the thickness of each layer and the production tolerance. This error is roughly in the 2$\sigma$ range. However, the overall performance is important, and both the 90/150~GHz and 220/280~GHz samples were well constructed.
\Add{For the 90/150~GHz sample, another AR coating using other dielectrics was available, and the performance of our coating was comparable to that of it.}

\begin{table}[tp]
\centering
\caption{Average reflectivity (on-axis performance) of the sample in each band. Differences between samples and designs mainly come from the thickness of coating materials. These errors are largely due to tolerance production. It is important that the overall performances are reasonable.}
\label{tab:relfectivity}
\begin{tabular}{c|cc|cc} \hline
       & \multicolumn{2}{|c|}{90/150~GHz} & \multicolumn{2}{|c}{220/280~GHz} \\ \cline{2-5}
       & 90 & 150 & 220 & 280 \\ \hline
measured & 1.7\% & 3.4\% & 0.38\% & 0.87\% \\ \hline
design   & 3.6\% & 0.9\% & 0.92\% & 0.18\% \\ \hline
\end{tabular}
\end{table}

\Erase{We also measured}The transmission to estimate the AR coating's loss tangent, tan$\delta$, \Add{was also measured}. By measuring the decrease in transmittance and combining it with the reflectance measurement, tan$\delta$ \Add{could}\Erase{can} be determined. The loss tangents of materials increase as the frequency increases, and it is necessary to estimate them, especially for the 220/280~GHz sample. At lower temperatures, the loss tangent is expected to decrease. Therefore, \Erase{we can place}an upper limit on the loss tangent \Add{could be added} by measuring the transmittance at room temperature.

\begin{figure}[tbp]
\begin{minipage}[t]{0.5\linewidth}
\centering
  \includegraphics[keepaspectratio, scale=0.18]{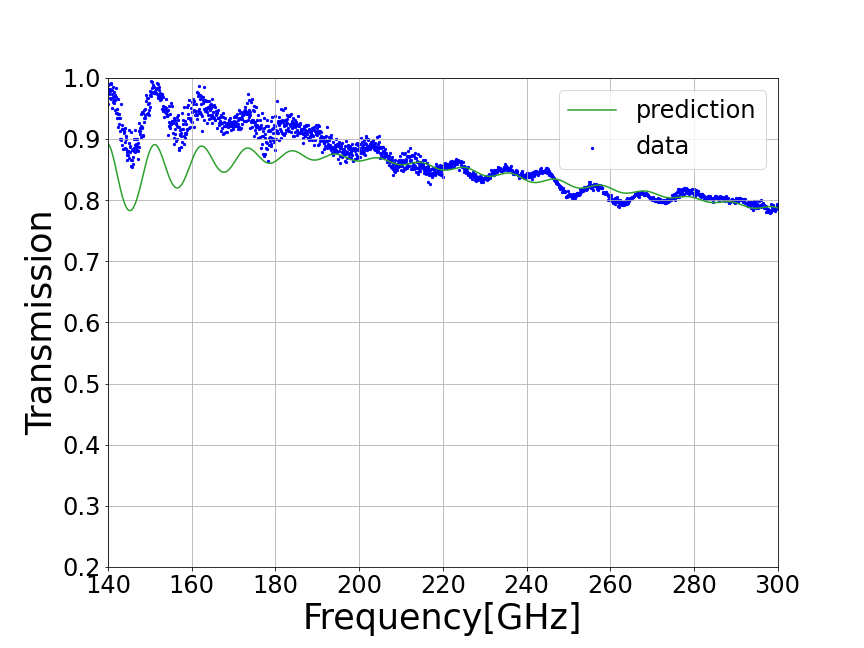}
\label{fig:fit UHF result}
\end{minipage}
\begin{minipage}[t]{0.5\linewidth}
\centering
  \includegraphics[keepaspectratio, scale=0.18]{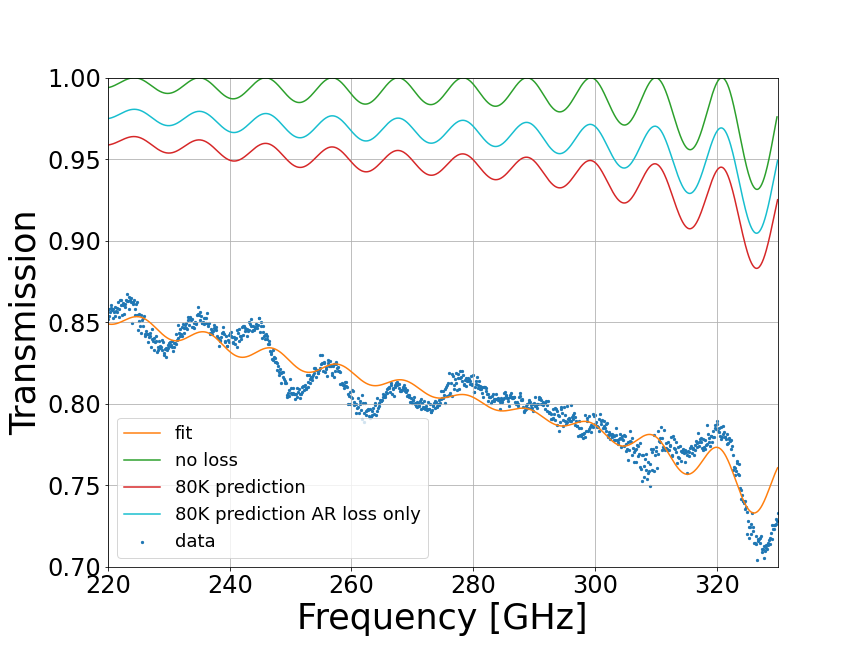}
\label{fig:trans UHF discussion}
\end{minipage}
\caption{Left: Measured transmission of the 220/280~GHz sample. For prediction, we substituted the index and thickness fit values obtained from the reflectance measurement. \Erase{Also, \Erase{we put}the fitted values (220-330 GHz only) \Add{represent the} loss tangent.}The fringe patterns of reflectance and transmittance were consistent. Right: Fitted transmission of the sample (orange) compared to the prediction of 80 K. \Erase{\Erase{We fixed}The indices and thicknesses \Add{were fixed} to the fit results in the reflection and fitted}\Add{Only loss tangents were fitted.} If no loss occurred, which would have been the ideal situation, transmission is shown as the green line. The 80~K prediction is shown as the red line. To evaluate the effects of coating materials, the cyan line represents the loss tangent of alumina\Add{, which was} set to 0 at 80 K prediction. The transmission increased $\sim$20\% at 80 K.}
\label{fig:transUHF}
\end{figure}

Figure \ref{fig:transUHF} shows the measured data of the 220/280~GHz sample. For prediction, the index and thickness fit values obtained from the reflectance measurement were substituted. First, it was confirmed that the fringe patterns of reflectance and transmittance were consistent, and the results showed that the measurement system was valid.
Then,\Erase{we fit} the materials' loss tangent \Add{was fitted} while fixing the indices and thickness to the values obtained from the reflectance measurement. After fitting the measured transmission, the transmission was estimated at about 80~K, \Add{the temperature} at which the HWPs and alumina filters operate. The AR coating’s loss tangent decreased at low temperature, and transmission \Add{was expected to} increase by $\sim$20\% at 80~K.
\Add{To estimate how much the average transmittance increases at the cryogenic temperature, the values of the alumina index were fixed according to Inoue et al. \cite{Inoue:16} and the other indices and thicknesses to the measured values at room temperature, and how much the average transmittance increased (how much absorption decreased) was estimated.}
Table \ref{tab:loss values} shows the fit values and prediction of the materials' loss tangents at 80~K.
\Add{Since these estimates rely on some of the values extrapolated from other frequencies or temperature, actual measurements must be taken to confirm them.}
The performance at low temperature will be evaluated in future work.

As a cryo-mechanical test, a cooldown test was also conducted to make sure that the fabricated samples \Add{did} not delaminate when cooled.
\Add{Duroid was diced into $4\times4$~cm square islands after the Epo-Tek has cured while the mullite layer was not diced} to prevent peeling due to heat shrinkage. \Erase{We cooled}The \Add{large diameter} sample \Add{was cooled} three times to 30~K at \Add{the} High Energy Accelerator Research Organization.
\Erase{There was}No delamination after cooldown \Add{occurred,} and the optical performance was the same as before cooling.
\Add{This cooldown test was performed every time a large sample was fabricated.}
\Add{After the cooldown test, an optical test was performed to confirm that the optical performance was unchanged compared to the earlier stage, indicating that no delamination had occurred. For optical measurements, a total of nine locations were measured (center and eight peripheral locations) to ensure that the coating was uniform. The detail of the verification process will be described in our future manuscript.}

\begin{table}[tb]
 \caption{Loss tangent, tan$\delta$ ($\times10^{-4}$), of the coating materials at 300 K (fit values) and 80 K (prediction). \Add{Systematic errors at 300~K were estimated from the measurement of the reference sample, alumina slab.} At lower temperatures, the loss tangent was expected to decrease, while the transmittance was expected to increase.}
  \centering
   \begin{tabular}{c|c|c|c} \hline
     Temperature & Alumina                           & Mullite                          & Duroid  \\ \hline
     300 K       & 7.0 $\pm$ 0.1                     & 423 $\pm$ 3                      & 39 $\pm$ 15  \\ \hline
     80 K        & 3.0 $\pm$ 1.1 \cite{Inoue:16} & 53 $\pm$ 10 \cite{Inoue:16} & $<$ 21 \cite{ROGERS}   \\ \hline
   \end{tabular}
  \label{tab:loss values}
\end{table}

\section{Conclusion}
\Erase{We have presented}The design\Add{s}, fabrication, and optical performance of a 2-layer AR coating for CMB polarimetry at 90/150~GHz and 220/280~GHz \Add{are presented}. Our coatings with mullite and Duroid 5880LZ \Add{led to a reduction in} the reflectivity to about 2.7\% at 90/150~GHz and $<1$\% at 220/280~GHz.
A HWP and an IR filter with our 90/150~GHz coating was used in the Simons Array experiment. This technique could also be applied to other CMB experiments, \Add{such as} \Erase{like}the Simons Observatory.

\begin{acknowledgements}
We would like to thank Junji Yumoto and Kuniaki Konishi for the X-ray CT measurement understanding thicknesses of small samples.
This work was supported by JSPS Core-to-Core program grant number JPJSCCA20200003, World Premier International Research Center Initiative (WPI), MEXT, Japan, and JSPS KAKENHI Grant Number 19H00674 and 19K14732. This research was supported by FoPM, WINGS Program, the University of Tokyo.
\end{acknowledgements}

%
%

\bibliographystyle{spphys}       
\bibliography{reference}   

\end{document}